\documentclass[epj,amsmath,amssymb]{svjour}

\usepackage{graphics,epsfig}
\usepackage{amssymb} 

\newcommand{\bi}[1]{\bibitem{#1}}
\newcommand{\lab}[1]{\label{#1}}

\newcommand{\ba}{\begin{eqnarray}}
\newcommand{\ea}{\end{eqnarray}}
\newcommand{\beqs}{\begin{eqnarray}}
\newcommand{\eeqs}{\end{eqnarray}}

\newcommand{\dd}{\Delta}

\begin{document}
\title{ New properties of  elastic $pp$ and $p\bar{p}$ scattering at high energies }



 \author{
O. V. Selyugin\fnmsep\thanks{\email{selugin@theor.jinr.ru}  }
}

\institute{ 
 BLTP,
Joint Institute for Nuclear Research,
141980 Dubna, Moscow region, Russia }


\abstract{ Data-driving determination of the new properties of elastic scattering  at small angles
  on the basis
  on  all existing experimental data  for $d\sigma/dt$ of  $pp$ and $p\bar{p}$
  at $\sqrt{s} \geq 540$ GeV
  allows us to obtain the main characteristics of the nonstandard terms of the elastic scattering amplitude.
  The energy dependence
  of the oscillation term  and the term with an  extremely large slope is determined.
  It was shown that  part of the oscillation term has a different sign for $pp$ and $p\bar{p}$ reactions;
  hence, it is  part of the Odderon amplitude.
  The period of the oscillation term  agrees with the scaling properties predicted by the Auberson -
    Kinoshita - Martin (AKM) theorem.
    The high quality quantitative description of  all data at $\sqrt{s} \geq 540$ GeV in the framework
    of the  high energy general structure (HEGS)    model supports such  a phenomenon which can be connected with peripheral hadron interaction.
  }


\maketitle


\section{Introduction}


  Now  many physical researchers  concentrate on  the search for new physics beyond the Standard Model.
   However, there are  
   such problems as confinement, hadron interaction at larger distances, non-perturbative hadron structure
   (parton distribution functions (PDFs), generalized parton distributions (GPDs) and others)
 that    should be explored in the framework of the Standard Model.
   These problems are connected with the hadron interaction at high and super-high energies
   and with the problem of  energy dependence  
   of the scattering amplitude and  total cross sections.
   This reflects   a tight connection of the main properties of  elastic hadron scattering
   with the first principles of quantum field theory   \cite{Block-85,royt}
   and the concept of the scattering amplitude as a unified analytic function of its kinematic variables
   \cite{Bogolyubov:1983gp}.

Researches into the structure of the elastic hadron scattering amplitude
   at superhigh energies and small momentum transfer, $t$,
      outline a connection   between
    experimental knowledge and  the  fundamental 
     asymptotic theorems,
   which are based on the first principles. 
   This connection provides  
    information about  hadron interaction
   at large distances where the perturbative QCD does not work \cite{Drem1},
       which gives the premise for a new theory to be developed.

   For example, it was shown  \cite{akm} that
   if the Pomeranchuk
   theorem was broken and the scattering amplitude
   grew to the maximal possible extent but did not break the Froissart boundary,
   many zeros in the scattering amplitude
   should be  present 
   in the nearest domain of
   $t \rightarrow 0$ in the limit $s \rightarrow \infty$.
      The search for some oscillations in  elastic scattering at small $t$ has a long story.
      For example, in  \cite{zar}, using the complex Regge poles, it was  shown
      that the peripheral contributions of inelastic diffraction processes lead to the appearence
      in elastic cross sections of large and small periodical structures on transfer momenta.
      In \cite{barsh}, a bump structure was obtained
      at small $t$. Many attempts were made to research  oscillations in the differential cross sections
      (for example \cite{kontr}).
   Really, with increasing energy   some new effects \cite{CS-PRL}
   in  differential cross sections  can be discovered at small $t$ \cite{L-range}.

  In the  papers \cite{selh95,gnsosc}, it was shown that
  AKM oscillations with scaling properties could exist
  in high-precision experimental data of the UA4/4   Collaboration.
   This  was confirmed now in \cite{Per-1}. An attempt to find such oscillations with high frequency  in the new
   experimental data, obtained by the ATLAS Collaboration, show practically zero result.
  However, the picture can be more complicated and especially it  
  concerns an energy dependence of such a periodical structure.

  There are many models that describe the elastic scattering at different energies. Two large reviews
  \cite{Jenk-rev,Pakanoni} \\
  present some of them. Note, the most famous models \cite{Soffer-Wu,Bourrely-14} and \cite{Kohara}.
  Some comparisons of our model with others are presented in \cite{HEGS1}.
  One of the important differences of our approach from other consists in  the different method
   of the calculations of $\chi^2$, which shows the difference between the experimental data and model calculations
  (see Appendix C). Also, in most part the model examines a large amount of experimental data, including
  the data on the diffraction dip-bump structure and  large momentum transfers.
  However, they do not take into account the Coulomb-hadron interference region (CNI),
  (for example, model \cite{Martynov} with 43 fitted parameters).  On the contrary,
   some models
   take into account the CNI region
  but do not take into account the diffraction dip-bump structure
  as, for example, the TOTEM Collaboration and \cite{Petrov-Tkach}.
  This leads to the construction of an artificial scattering amplitude,
   that does not work at more large momentum transfer.

   Now we examine  some peculiarities of the scattering amplitude assuming the existence
  of the potential of  hadron-hadron interactions at large distance
    and carry out a new high accuracy  treatment of all experimental data  at $\sqrt{s} \geq 540$ GeV.

   The new data obtained at the large hadron collider (LHC) by
    the TOTEM and ATLAS Collaborations  made it possible to accurately analyse
     our model assumptions.
    Of course, to analyze new data more accurately
   in order  to find some new peculiarity in  hadron elastic scattering,
     one needs to have a simple model as a background which,
    with  a few free 
     parameters,  gives 
     a good description of specific properties such as the form and energy dependence of the diffraction minimum-maximum
       and description of the Coulomb-hadron region   of experimental data.
    The high energy generalized structure (HEGS) model \cite{HEGS0,HEGS1}
    was chosen for our analysis (see Appendices A and B).
    The model is based on the general quantum field theory principles (analyticity, unitarity, etc.) 
    and takes into account the basic information on the structure of a nucleon as a compound system and
    the hadron structure at large distances through the generalized parton distribution function \cite{Muller,Ji97,R97,Diehl}.
    The obtained GPDs \cite{GPD-ST-PRD09,GPD-PRD14}  allow one to simultaneously
      calculate the electromagnetic and gravitomagnetic form factors
   from one     form of GPDs.
     The HEGS model uses both these
    form factors without any free parameters, 
    which provides  the hard structure of the model.
    In the framework of the model, a good quantitative description of the $pp$ and $p\bar{p}$
    was obtained \cite{HEGS1}
   ($3416$ experimental points were included in the analysis
 in the energy region   $9.8$ GeV $\leq \sqrt{s} \leq 8.$ TeV
 and in the region of momentum transfer $0.000375 \leq |t| \leq 15 $ GeV$^2$).
 The experimental data 
 are included
 in 92 separate sets of 32 experiments \cite{data-Sp,Land-Bron}.
  The whole Coulomb-hadron interference region,
  where the experimental errors are remarkably small,
    was included in our data analysis. 
 As a result, it was obtained for the number of experimental data $N=3416$
   $ \sum_{i=1}^{N} \chi_{i}^{2}/N=1.28$
   with only $8$ fitting parameters.

\begin{table}
\label{Table_1}
\caption{ Sets of $\frac{d\sigma}{dt}$ 
($k_i$ -additional normalization coefficient).
  }
\vspace{.5cm}
\begin{tabular}{|c|c|c|c|c|c|c|} \hline
re.   &$\sqrt{s}$& Exp.& $n$ &$-t_{min}10^{3}$ & $-t_{max}$ & $k_{j}$  \\ 
           & TeV &        &    &$(GeV^2)$        & $(GeV^2)$  &                        \\ \hline
$pp$       &13   & \cite{ATLAS-13}   &79    & 0.29     & 0.4376     & $ 1.09 $  \\ 
$pp$       &13   & \cite{70-T8c}  &138   & 0.88     & 0.20    & $ 0.97 $  \\ 
$pp$       &13   & \cite{TOTEM-13a}   &273   & 43.3     & 1.4     & $ 0.97 $  \\ 
$pp$       & 8   & \cite{TOTEM-13b}    &42    & 201.86   & 1.76     & $ 0.98 $  \\ 
$pp$       & 8   & \cite{64-T8}   &39    & 10.5     & 0.36     & $ 1.09 $   \\ 
$pp$       & 8   & \cite{65-T8b}   &31    & 0.741    & 0.20      & $ 0.98 $  \\ 
$pp$       & 8   & \cite{TOTEM-8nexp}   &30    & 28.5     & 0.2     & $ 1.0  $  \\ 
 $pp$      & 7   & \cite{63-ATLAS-8}   &40    & 6.2      & 0.36     & $ 1.08 $  \\ 
$pp$       & 7d  & \cite{TOTEM-8nexp}   &43    & 5.15     & 0.13       & $ 1.02 $  \\ 
$pp$       & 7d  & \cite{T7a}   &60    & 474.     & 2.44      & $ 1.04 $  \\ 
$pp$       & 2.76& \cite{T2p76}   &45    & 72.46    & 0.46     & $ 0.93 $   \\ 
$pp$       & 2.76& \cite{T2p76}   &18    & 371.9    & 0.74     & $ 0.91 $  \\ 
$p\bar{p}$ & 1.96& \cite{D0-1p96}    &16    & 260.     & 1.2        & $ 1.05 $  \\ 
$p\bar{p}$ & 1.8 & \cite{data-Sp}     &10    & 925.     & 1.4      & $ 0.77 $   \\ 
$p\bar{p}$ & 1.8 & \cite{E710-1p8}   &50    & 33.9     &0.63       & $ 1.11 $  \\ 
$p\bar{p}$ & 1.8 & \cite{CDF-1p8a}     &25    & 35.      &0.29       & $ 0.97 $  \\ 
$p\bar{p}$ & 0.63& \cite{Bernard86}    & 19   & 730.     & 2.25       & $ 0.98 $  \\ 
$p\bar{p}$ & 0.54&\cite{UA4/2}  &97    & 0.875    &0.1162      & $ 1.13 $  \\ 
$p\bar{p}$ & 0.55& \cite{data-Sp}   &66    &  2.25    & 0.035    & $ 1.06 $   \\ 
$p\bar{p}$ & 0.55& \cite{data-Sp}    &14    & 26.0     & 0.08      & $ 1.07 $  \\ 
$p\bar{p}$ & 0.55&  \cite{UA4-87-21}     &58    & 32.5     & 0.32     & $ 1.04 $  \\ 
$p\bar{p}$ & 0.55&  \cite{UA4-87-22}     &29    & 215.     & 0.5      & $ 0.99 $  \\ 
$p\bar{p}$ & 0.55&  \cite{UA4-87-23}     &33    & 460.     & 1.53       & $ 1.01 $  \\ 
$p\bar{p}$ & 0.54&  \cite{data-Sp}     &33    & 75.      & 0.45      & $ 1.12 $  \\ 
   \hline     \hline
 & $\sum $ & &1288    &    &   &      $\overline{1.015}$    \\ 
    \hline
\end{tabular}
\end{table}

   The high precision experimental data obtained by the TOTEM and ATLAS Collaboration at
   $\sqrt{s} = 7 - 13$ TeV show a significant deviation from the standard exponential behavior
   at small momentum transfer \cite{Khoze-Sl}, as first observed by the TOTEM Collaboration \cite{T7a,64-T8,TOTEM13-1set,TOTEM13-2set}.
   Regardless of the concrete model,
   some tension between the data of the TOTEM and ATLAS Collaborations is observed,
   which is reflected in the  definition 
   of the sizes of the total cross sections.
   The usual practice of taking into account  the fitting procedure of the experimental errors
    in quadrature of statistical and systematical errors does not remove this tension but only decreases
    the value of $\chi^2$.  As a result, in some papers the analysis was carried out
    for the  TOTEM and ATLAS data separately, for example  \cite{L-P-22}.
    In contrast,
    our strategy in all our model analysis is vastly different (see Appendix C).
     In the standard fit, the statistical errors
    are taken into account. 
    Since an uncertainty of luminosity   gives the main contribution to
    the systematic errors,  it is taken into account in the form of an additional normalization
    coefficient, which is equal for all data of one set. As a result,  all data of all sets
    of  different experimental Collaborations are  taken into account simultaneously.

       Using  our model for the analysis of the new data of the TOTEM Collaboration at $\sqrt{s}=13$ TeV
    shows the existence of  new peculiarities in the differential cross section at small angles.
    It was shown that the differential cross section has some small oscillatory behavior \cite{osc-13}.
     This effect     
      was confirmed independently of a concrete model by the statistically based method \cite{Hud,osc-13}  also at a high confidence level  (see Appendix D).
      Earlier, the oscillation behavior of the differential cross section 
      was observed in the data of the   UA4/2 Collaboration at the SPS accelerator at $\sqrt{s}= 541$ GeV \cite{gnsosc}
    and at low energies \cite{osc-3p7}.

 Further careful analysis also showed the existence in the scattering amplitude
   of  the term with an extremely large slope \cite{fd13}.
  In the work \cite{fd13}, the analysis of both sets of the TOTEM data at 13 TeV
     was carried out with  additional normalization equal to unity and taking into account
     only  statistical errors in experimental data.

\begin{figure}
%
 \vspace{-0.5cm}
\includegraphics[width=0.5\textwidth]{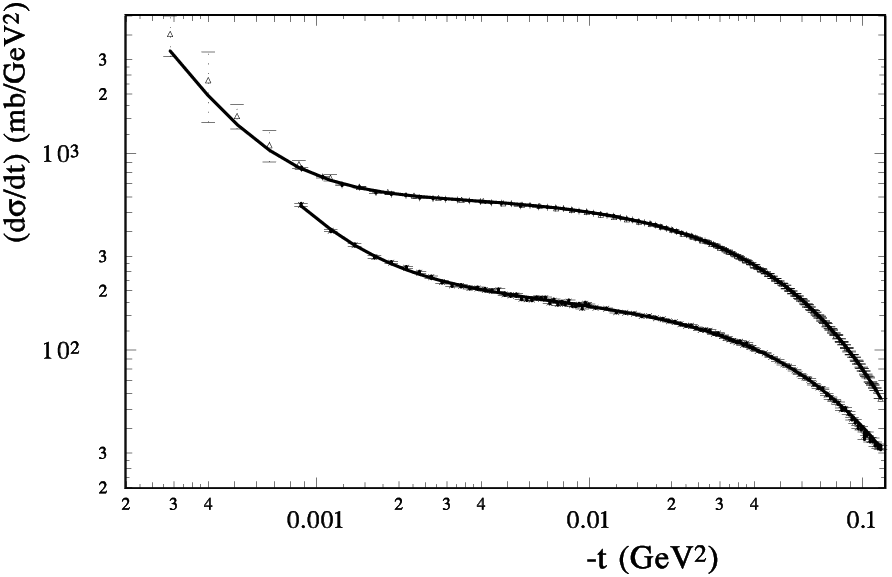}
\vspace{1.5cm}
\caption{The differential cross sections are calculated in the framework of the HEGS model
 with the additional terms  for $pp$ scattering at $\sqrt{s} = 13$ TeV (upper line and data)
 and  for $p\bar{p}$ scattering  at $\sqrt{s} = 541$ GeV (lower line and data).
  }
\label{Fig_1}
\end{figure}

\begin{figure}
%
 \includegraphics[width=0.5\textwidth]{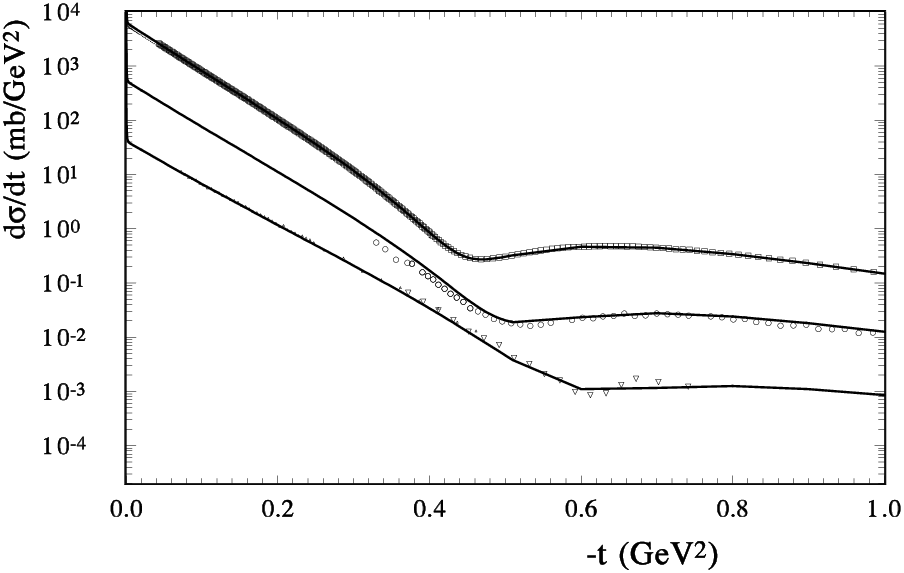}
\vspace{1.5cm}
\caption{The differential cross sections are calculated in the framework of the HEGS model
with the additional terms (anomalous and oscillation)
at $\sqrt{s} = 13$ TeV (upper line and data) multiplied by $10$; at  $\sqrt{s} = 7$ TeV
and at  $\sqrt{s} = 2.76$ TeV \cite{T2p76} divided by $10$ (lower curve and data).
  }
\label{Fig_2}
\end{figure}

    Both effects have been confirmed in the analysis of
    all sets of the experimental data obtained at the LHC \cite{fd-LHC}.
    However, to study the properties of the new effects in  elastic scattering, one should
    include in the analysis the experimental data for proton-antiproton scattering and essentially extend
    the energy region.

%

\begin{figure}[t]
  \vspace{-0.5cm}
\includegraphics[width=0.5\textwidth]{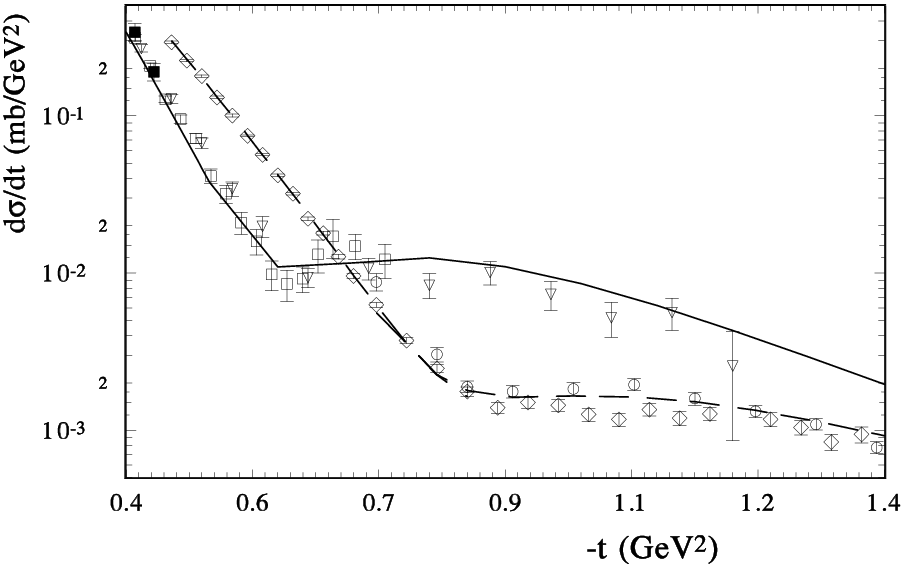} 
\vspace{1.5cm}
\caption{The $d\sigma/dt$ of  elastic $pp$ scattering
(calculated in the model with Odderon contribution and two additional terms)
 for $pp$ scattering at  $2.76 $ TeV (the data of the TOTEM Collaboration - squares)
 are  compared with the calculations and data for $p\bar{p}$ at $\sqrt{s}=1.8$ TeV ( triangles down) and
 at $\sqrt{s}=546$ GeV (dashed line and diamond and circles).
  }
\label{Fig_3}
\end{figure}

\section{$s$ and $t$ dependence of new peculiarities in elastic scattering}
     In this work,  all high energy  data of  $pp$ and $p\bar{p}$  scattering are
      included in the analysis.
    Especially, this is important in view of a  wide discussion  \cite{Her} of the comparison of the Odderon contributions  to  $p\bar{p}$ scattering of the D0 experiment at $\sqrt{s} = 1.96$ TeV  \cite{D0-1p96} and
    to $pp$  scattering of
    the TOTEM experiment  at $\sqrt{s} = 2.76$ TeV \cite{T2p76}.
   Also,   the high precision data of the UA4/2 Collaboration related to
    $p\bar{p}$ scattering should be taken into account.
     Hence, let us include in our analysis all experimental data at
     $\sqrt{s} \geq 540$ GeV  (see Table 1).

  As was found from an analysis of the TOTEM data at $13$ TeV, there exists an additional anomalous term
  in the scattering amplitude with an essentially larger slope than the standard term \cite{fd13}.
  Usually, some terms with a large slope are related to the second Reggions.
   However, in all models the second Regge terms (like $f_0, \rho, \omega ...$) show a decrease
   in their contributions with energy, 
   like $1/s^{1/2}$.    As a result of our analysis,
   the form of an additional anomalous term was determined     as
\ba
 f_{an}(t)= && i h_{an} \ln{(\hat{s}/s_{0})}/k  \\ \nonumber
  && \exp[-\alpha_{an} (|t|+(2t)^2/t_{n})\ln{(\hat{s}/s_{0})}] \ F_{em}^2(t);
\ea
where $h_{an}$ is the constant determining the size of the anomalous term with a large slope - $ \alpha_{an}$;
 $  F_{em}(t)$  is the electromagnetic form factor,
  which was determined 
   from the 
  GPDs \cite{GPD-PRD14}, and
  $ \ k=\ln(13000^2  \ {\rm GeV}^{2}/s_{0}) \ $  is  introduced for normalization of $h_{an}$ at $13$ TeV,
  $t_{n}=1$ GeV$^2$-normalization factor (see Appendix B for definitions of $\hat{s}$ and $s_{0}$).
 Such a form adds only two additional fitting parameters,
 and
   this term is supposed to grow with energy of order $ \ln{(\hat{s}/s_{0})}$.  The term has a large imaginary part
and a small real part determined by the complex $\hat{s}$.
  Note that for analysis of the basic properties of elastic scattering,  
  the experimental data are usually taken
  in a narrow region of momentum transfer.
  This  can lead to a bad description if we take a wider region.
  Now all terms of our scattering amplitude are used in all regions of the  available experimental data.

 Now let us try to find the form of an  additional oscillation contribution
   to the basic elastic scattering amplitude (see Appendix B).
   Our fitting procedure takes
  the oscillatory function
\vspace{-0.1 cm} 

\ba
 f_{osc}(t)= &&  i  h_{osc} (1 \pm i) \ln{(\hat{s}/s_{0})}/k   \   J_{1}(\tau)/\tau \ A^{2}(t), \\
  && \tau = \pi \ (\phi_{0}-t)/t_{0}; \nonumber
\ea
here $J_{1}(\tau)$ is the Bessel function of the first order;
     $t_{0}=1/[a_{p}/(\ln{(\hat{s}/s_{0})}/k)]$, where $a_{p}=17.15$ GeV$^{-2}$ is the fitting parameter,
     that leads to AKM scaling on  $\ln{(\hat{s}/s_0)} $;
     $A(t)$ is gravitomagnetic form factor, which was determined from the
  GPDs \cite{GPD-PRD14} and
   $h_{osc}$ is the constant that determines the amplitude of the oscillatory term with
   the period determined by $\tau$.
 This form has only a few additional fitting parameters and allows one to represent
 a wide range of  possible oscillation functions.
 For simplicity, the phase $\phi_{0}$ is taken as zero
  for $pp$  and small value for  $p\bar{p}$ scattering.
 Inclusion in the fitting procedure of the data of  $p\bar{p}$ elastic scattering shows
 that the part of oscillation function changes  its sign for the crossing reactions.
 As a result, the plus sign is related with  $pp$ and minus with  $p\bar{p}$ elastic scattering.
   Hence, this part is  the crossing-odd  amplitude, which has the same simple form for $pp$ and
  $p\bar{p}$ scattering only with different sign.  Of course, it's analytical properties require further researches.

   The wider energy region used in this analysis allows one to reveal the logarithmic energy dependence of the
   oscillation term.  Let us compare the constant (size) of the oscillation function of three
   independent analyses  (only $13$ TeV, all LHC data, all data above $500$ GeV ): 
   $  h_{osc}^{a}=0.350\pm0.014 \ {\rm GeV}^{-2}; \ h_{osc}^{b}=0.370\pm0.013 \  {\rm GeV}^{-2}; h_{osc}^{c}=0.270\pm0.007 \ {\rm GeV}^{-2}$.
   The size of $  h_{osc}$ is smaller in the last case; however, the error is decreased.
   Perhaps,  this reflects a more complicated form of  energy dependence, for example, we can not
   determine the power of $\ln(s)$ exactly and take it as unity.
   In future, we intend to study this problem using an essentially  wider energy interval
   up to $\sqrt{s}=3.6$ GeV, where some oscillation  was also observed in $p\bar{p}$ scattering \cite{osc-3p7}.
    Note, despite  the logarithmic growth of the oscillation term, its relative contribution  decreases
    as the main scattering amplitude grows as $\ln^{2}(s)$.

\begin{figure}[b]
 \includegraphics[width=0.5\textwidth]{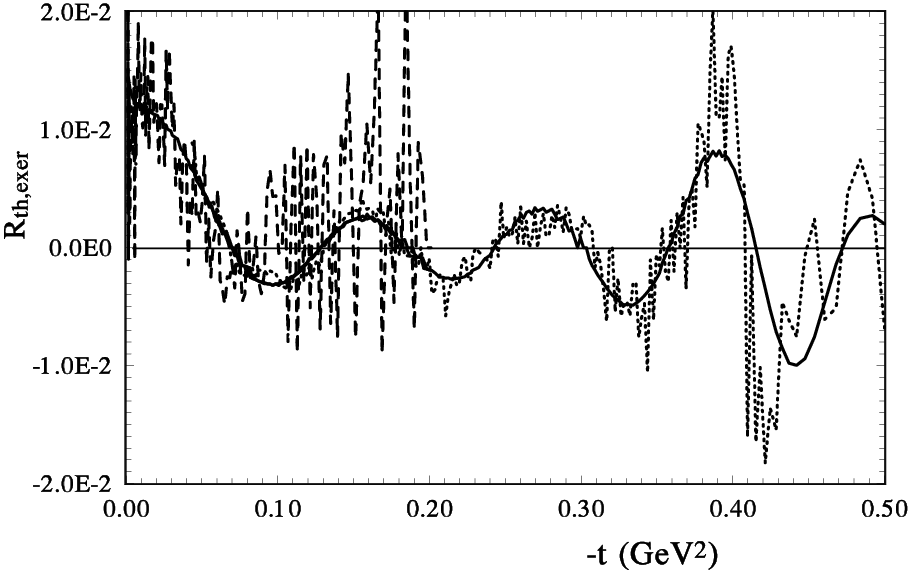}
\vspace{1.cm}
\caption{  $R_{\mathrm th} $ of eq. 3  (the solid thick line)  and $R_{\mathrm exper.} $ of eq. 4
  of the TOTEM data at  $\sqrt(s)=13 $ TeV  (the dashed line). 
%
  }
\label{Fig_4}
\end{figure}

\begin{figure}[b]
\includegraphics[width=0.5\textwidth]{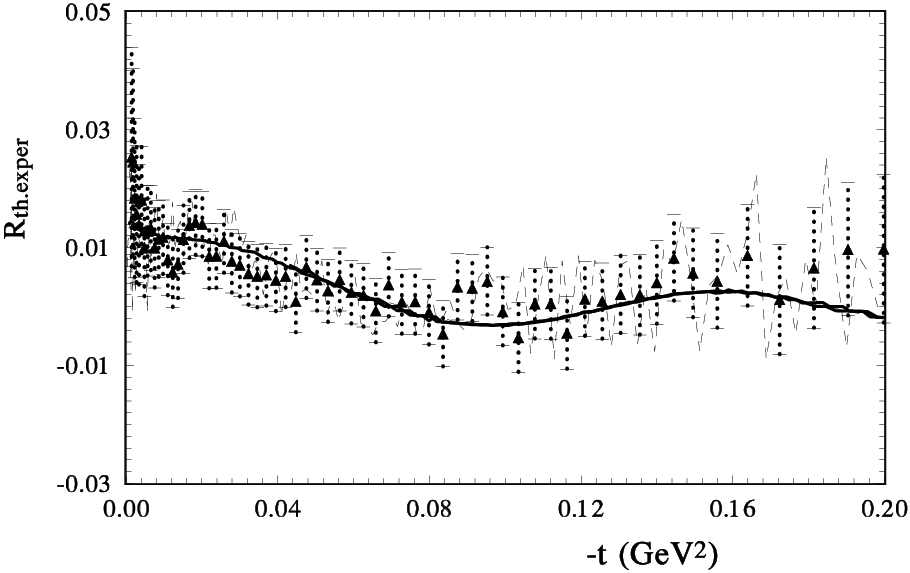}
\vspace{1.cm}
\caption{ $R_{\mathrm th}$ of eq.3   (the solid thick line) and $R_{\mathrm exper.} $ of eq.(4)
   (the triangles up) at $\sqrt(s)=13 $ TeV (ATLAS data).
  }
\label{Fig_5}
\end{figure}

 After the fitting procedure, with the modern version of FUMILY \cite{Sitnik1,Sitnik2},
  we obtain  $\chi^2/dof =1.23$ (remember that we used only statistical errors).
 The total number of  experimental points
 of $24$ sets equals $1288$ and the sum of $\sum \chi^2 = 1568$.  If we remove the oscillatory function, then
 $\sum \chi^2 = 4344$   increases significantly. 
  If we make a new fit without $f_{osc}$,
 then $\sum \chi^2 = 2512$ decreases but remains large.
 However, the basic parameters change slightly
 (see second column below);

\begin{table}
\label{Table_2}
\caption{Comparisons of the values of the constants (eq. 11 and eq. 12) and (eq. 1 and eq. 2) are obtained by
  the fitting procedure with
  and without the oscillation term (eq.2).
  }
\vspace{.5cm}
\begin{tabular}{|c|c|c|} \hline
  & with  the oscillation &  without the oscillation   \\
           &term   &     term                          \\ \hline
$C_{\mathbb{P}}$       &$3.30  \pm  0.02$    & $3.36  \pm  0.02 $   \\ 
$C'_{\mathbb{P}}$      &$1.39  \pm  0.02$   & $1.35  \pm  0.02  $      \\ 
$ C'_{\mathbb{O}}$     & $-0.56 \pm 0.03$   & $ -0.56 \pm 0.03 $   \\ 
$h_{an}$               & $ 2.17  \pm 0.03$  & $ 1.94  \pm 0.03 $   \\ 
$\alpha_{an}$          & $0.51  \pm 0.01 $  & $ 0.56 \pm 0.01  $     \\ 
   \hline
\end{tabular}
\end{table}






 Our model calculations  for the differential cross sections
 are represented
  in Fig. 1, Fig. 2  and Fig. 3.
   It can be seen that the model gives a beautiful description of the differential
  cross section especially at LHC energies in both  the Coulomb-hadron interference region and  the region of the diffraction dip for all energies.
 A phenomenological form of the scattering amplitude  determined for small $t$ can lead
   to  very different differential cross sections at larger $t$.
  This effect is  especially prominent when it is connected with the differential cross section at 13 TeV,
   as the diffraction minimum is located  at a non-large $t$.
  An important point of obtaining a good 
  description of  the available experimental data in a wide region of energies and momentum transfer is that
     two additional specifical functions
     are used in the model  in the whole region of Mandelstam's variables. 

 As a result,
   two parameters of the additional  term of eq. (1)  are well defined
   $ h_{an}=2.10 \pm 0.05 \ {\rm GeV}^{-2};  \ \ \ \alpha_{an} =0.51 \pm0.03 \ {\rm GeV}^{-2} $.
  The sizes of constant $ h_{an}$ obtained in our three different
   studies  can be compared  for 3 different studies (first analysis \cite{fd13} include only $13$ TeV experimental data; the second analysis \cite{fd-LHC}  taken into account all data on elastic scattering  obtained at LHC; third case (present analysis) taken into account  all data above $500$ GeV, including proton-antiproton data). If it is an artifact,
  presented only at 13 TeV experiments, the size of $h_{an}$ has to be decreased when we include more and more experimental data. However we have for these 3 studies:
   $  h_{an}^{a}=1.70\pm0.05  \ {\rm GeV}^{-2} $;  $h_{an}^{b}=1.54 \pm 0.08 \ {\rm GeV}^{-2}$ ;
    $ h_{an}^{c}=2.10 \pm0.05 \ {\rm GeV}^{-2} $.
   The maximum value was obtained in the last case when we used the experimental data in a wider region of $s$ and $t$, including the data of  crossing reaction.

  To see the oscillations in the differential cross sections, let us
  determine two values - one is purely theoretical 
\ba
  R_{\mathrm th}(t) = \frac{d\sigma/dt_{\mathrm th0+osc}}{d\sigma/dt_{\mathrm th0}}-1,
\ea
and other with experimental data
\ba
  R_{\mathrm exper.}(t)= \frac{d\sigma/dt_{exper.}}{d\sigma/dt_{th0}}-1.
\ea
 Here $d\sigma/dt_{th0}$ is the model calculation without the oscillation part and
 $d\sigma/dt_{\mathrm th0+osc}$ is the model calculation with the oscillation part.
  The corresponding values are calculated from  our fit of all data and can be derived    
   from every separate set of experimental data.

\begin{figure}
\includegraphics[width=0.5\textwidth]{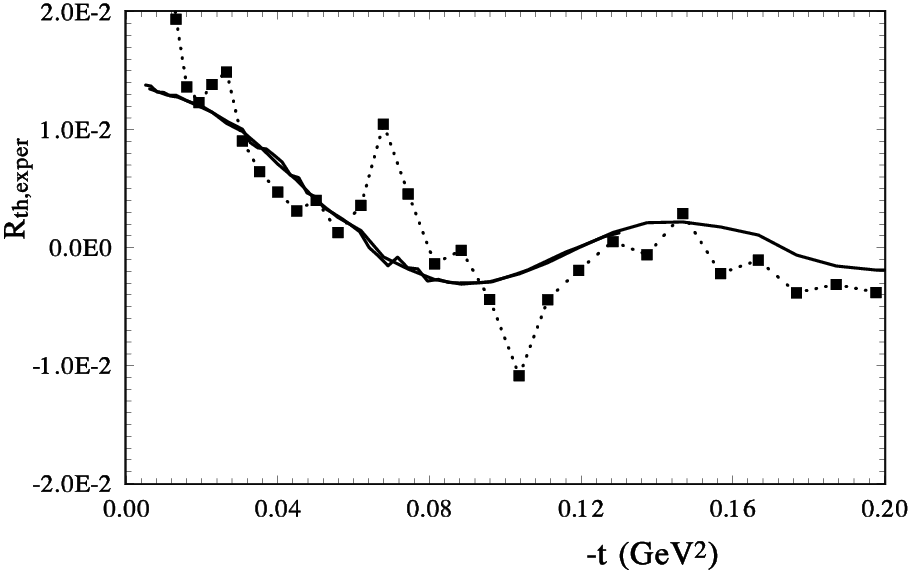} 
\vspace{1.cm}
\caption{ $R_{\mathrm th} $ (the solid thick line) and $R_{\mathrm exper.} $
   (the short dash  line and squares) are 
    the TOTEM data at $\sqrt(s)=7 $ TeV.
  }
\label{Fig_6}
\end{figure}

\begin{figure}
 \includegraphics[width=0.5\textwidth]{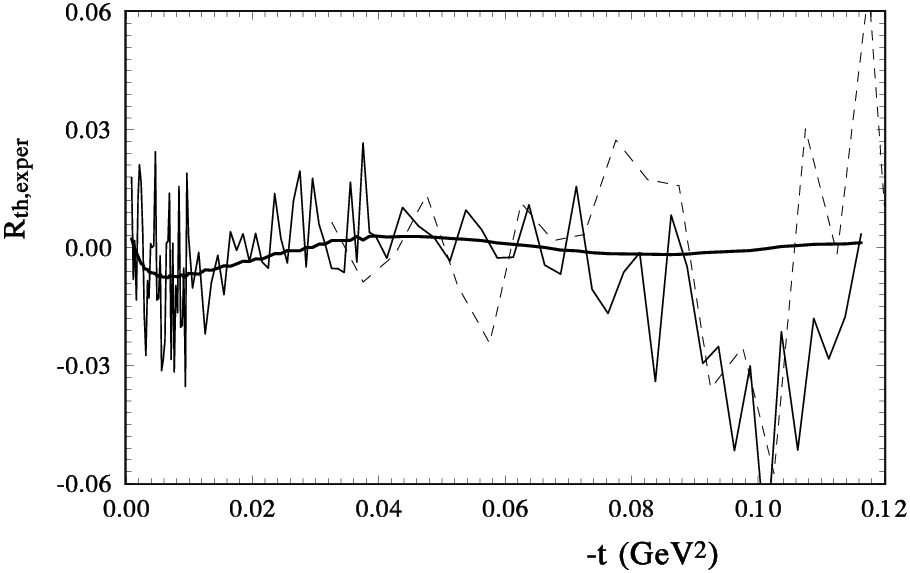}   
\vspace{1.cm}
\caption{ $R_{\mathrm th} $  (the solid thick line) and $R_{\mathrm exper.} $
   (the solid line, the data of the UA4/2 Collaboration and dashed line, the data of UA4 Collaboration)
 at  $\sqrt(s)=541$ GeV and $\sqrt(s)=546$ GeV.
  }
\label{Fig_7}
\end{figure}

 For two sets of the TOTEM and ATLAS data at $13$ TeV  the values $R_{\mathrm th}(t)$  and
  $ R_{\mathrm exper.}(t)$ are presented in Fig. 4 and Fig. 5.
 We can see that $R_{\mathrm th}$
    is similar to the value $R_{\mathrm exper.}$. The oscillation contribution is small;
     however, the noise of
    the background  decreases with $t$ and  does not damp
    the oscillation part.
    In Fig. 6, these values are presented for  experimental data obtained
    at $\sqrt{s}=7$ TeV.

\begin{figure}
\vspace{-1.cm}
\begin{center}
.\hspace{1cm} \includegraphics[width=0.5\textwidth]{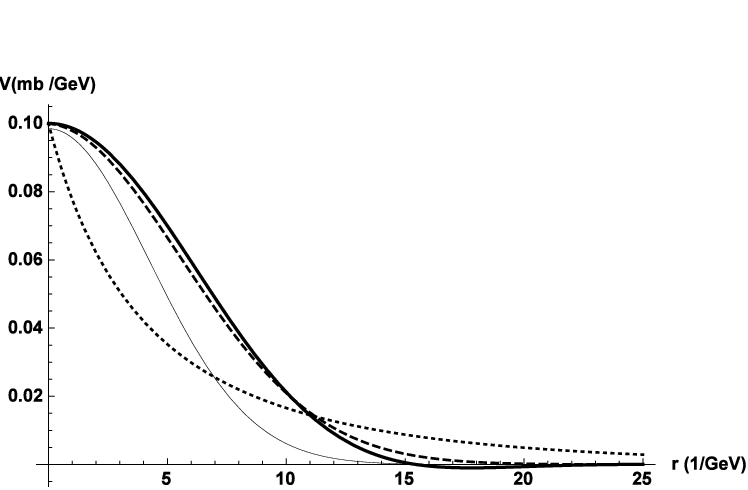} 
 \end{center}
\caption{ The different form of the peripheral potentials are compared.
   The modified Gaussian form of the potential $ h_{gs}\exp[\alpha_{gs}r^2]/(1 + r/r_{0})$ ( dashed line),
    the calculation of our anomalous term in the $r$-representation eq. (5)  (the solid line);
    for comparison,   the $r$-representation of the standard exponential behavior in the $q$-representation
 (thin solid line) and exponential behavior in $r$-representation
  $ h_{e}\exp[b_{e} r]/(1 + r/r_{0})$ (dotted line)
  are also shown.
  }
\label{Fig_8}
\end{figure}

 The corresponding values calculated from the fit of two sets of the UA4 and UA4/2 data at $541$ GeV are presented in Fig. 7.
  At small $t$, there is a large noise;  however, the oscillation
  contributions can be seen.
     Such a periodical structure has a sufficiently  long period, especially  compared to the
   periodical structure   with high frequency analysed in the previous work \cite{gnsosc,Per-1}.
   Our analysis of high frequency oscillations shows that they can occur 
   at $13$ GeV with a small amplitude,
   which means an essential    decrease
    compared to the oscillations at $541$ GeV. 
  The stability  of such amplitude is determined with a large
    error  $h_{osc}= 0.014 \pm 0.012$ GeV$^{-2}$. 

\begin{figure}
\begin{center}
.\hspace{1cm} \includegraphics[width=0.5\textwidth]{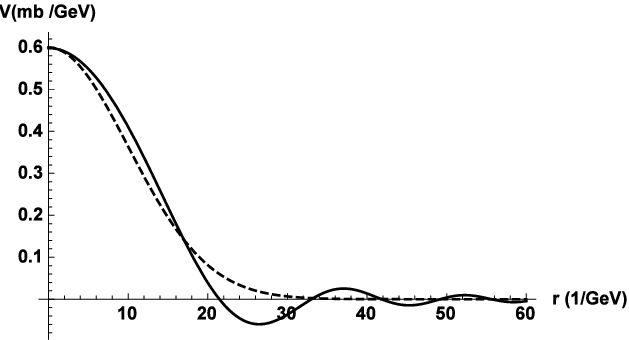}
 \end{center}
\caption{  The oscillation term eq. (2) in the $r$-representation eq. 6 (solid  line)
is compared with the Gaussian form $h_{g} \exp[b_{g} r^2]$ (dashed line).
  }
\label{Fig_9}
\end{figure}


\begin{figure}[b]
\includegraphics[width=0.5\textwidth]{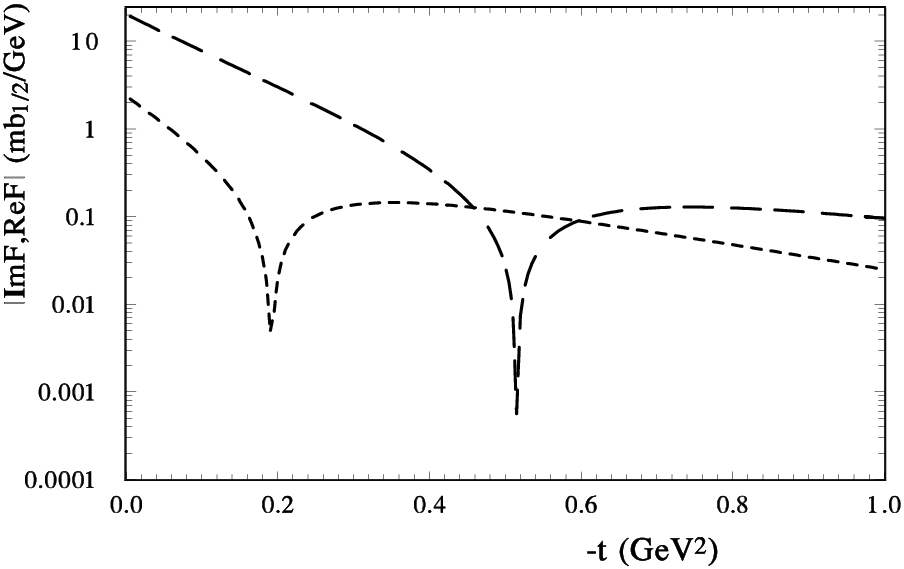}
\vspace{1.cm}
\caption{  The imaginary (long dashed line) and real (short dashed line) parts
 of the main (without oscillation term) elastic hadron
 scattering amplitude at 13 TeV.
  }
\label{Fig_10}
\end{figure}

   In Fig. 8,  different forms of  peripheral potentials are compared.
  The dotted line represents the modified exponential form of the potential 
  $ h_{e}\exp[b_{e} r]/(1 + r/r_{0})$,
   the solid line
    shows the calculation of our anomalous term in the $r$-representation
 \ba
V_{an}(r)=\int q \ dq \sin(q r)/r  \exp[-\alpha_{an} (q^2+(2q^2)^2/q^{2}_{0} ) ]. \lab{vfil}
 \ea
  It is compared with the modified Gaussian form of the potential (dashed line)
  $ h_{gs}\exp[\alpha_{gs}r^2]/(1 + r/r_{0})$,
where  $ h_{gs}$ is a constant providing the same size as our $V_{an}(r)$ (eq. 5) and  $r_{0}$ is a selectable interaction radius.
  Obviously, the last form well reproduces our anomalous term in  the $r$-representation.
 For comparison,   the $r$-representation of the standard exponential behavior in the $q$-representation
 (thin solid line) is also shown.

   The oscillation term, eq. (2), in the $r$-representation is presented  in Fig. 9 (solid thick line)
\ba
    V_{osc}(r) = \int  q \ dq \sin(q r)/r  J_{1}[\tau]/\tau, \lab{vfil}
 \ea
  where $J_{1}[\tau]/\tau$ is determined by eq. (2).
 It is compared with the Gaussian form $h_{g} \exp[b_{g} r^2]$ (dashed line)
 where  $ h_{g}$ is a constant, providing the same size as our $V_{osc}(r)$ (eq. 6) and $b_g$ is the fitting Gaussian radius.
 Of course, it does not reproduce
  small oscillations. Oscillations will appear in the $q$-representation if we cut the potential
 at a large distance.

\section{Some other  possible origins of periodic structure} 

The structure of the differential cross sections of elastic scattering has a complicated form
   that is dependent
  on $s$ and $t$ (see, for example, \cite{HEGS-min}).
  First the diffractive properties of elastic scattering are represented in the dip-bump structure
  that reflects the eikonalization of the Born scattering amplitude with the $s$ and $t$ dependence
  of its real and imaginary part. Of course, a periodic structure can be determined
  by the zeros of the real and imaginary parts of the scattering amplitude. Thus, the diffraction minimum is determined
  by the zero of the imaginary part. 
   The dispersion relations require that the real part
  of the amplitude has zero at small momentum transfer. However, the $s$ and $t$ dependence essentially differs from
  the periodic structure presented in the paper.
  The corresponding real and imaginary parts of the elastic scattering amplitude at 13 TeV are shown in Fig. 10.

\begin{figure}[t]
 \includegraphics[width=0.45\textwidth]{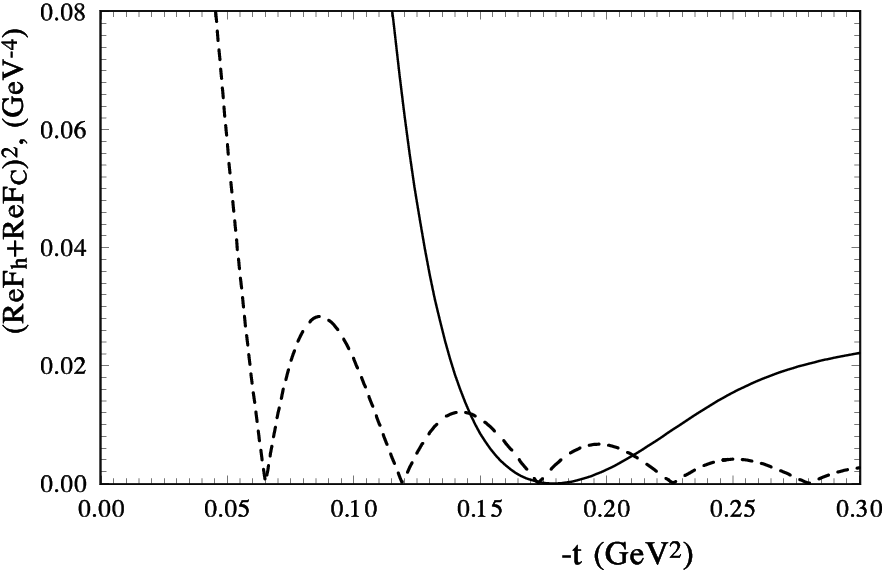} 
\vspace{1.cm}
\caption{The  comparison of $ \Delta_{R}^{th}(s,t)$, eq. (7) (solid line), of  elastic $pp$ scattering
 at $13$ TeV and  $|Re(f_{osc})| +  |Im(f_{osc})|$ of the oscillation function of eq. (2) (dashed line).
  }
\label{Fig_11}
\end{figure}

   There is an additional zero at some value of $t$
  where the real part of the Coulomb amplitude (which has the negative sign) equals in absolute value the real part of the  hadronic
  amplitude.  
  Let us determine the value $ \Delta_{R}$
that is dependent on the size of the real part of the scattering amplitude
\begin{eqnarray}
    \Delta_{R}^{th}(s,t)=(ReF_C(t)+ReF_h(s,t))^2.
    \label{Dr}
\end{eqnarray}
  Obviously, it gives the minimum at one point of $t_{min}$ where the real part of the Coulomb amplitude
   is opposite to the real part   of the scattering amplitude.
   the real part   of the scattering amplitude.
  Let us compare this value with our periodic structure, see Fig. 11.
  It is obvious that the $t$ dependence is very different for the examined values.

\section{Conclusion}
       The existence of the new effects   in high energy elastic $pp$ and $p\bar{p}$
    scattering is revealed by using the data-driving method for the first time at a  quantitative level.
      Using the HEGS model based on the GPDs allows us to describe at a  quantitative level
      all experimental data       on $pp$ and $p\bar{p}$ elastic scattering  above  $\sqrt{s} = 540$ GeV
      with taking into account only statistical errors.
      However, 
       only the presence 
        in the model of two anomalous terms, determined by the interaction
       in the  peripheral region 
        allows us to obtain a sufficiently small $\chi^2$.
             These additional terms affect the value of the total cross sections and parameter
             $\rho(s,t)$ - ratio of the real to imaginary part of the elastic scattering amplitude.
             As a result, we have obtained at $\sqrt{s} =13$ TeV
             $\sigma_{tot} = 110.4 \pm 0.4$ mb and $\rho(t=0) =0.106 \pm0.005$.

      The phenomenon of  oscillations  of the elastic scattering amplitude
       gives us  important information about
      the behavior of the hadron interaction potential at large distances.
       Earlier,  the existence of such oscillations in experimental data at $\sqrt{s}=13$ TeV was shown
      at the statistical  level by three methods: a) the method of  statistically independent selection;
      b) the comparison of the $\chi^2$ without oscillation ($\sum \chi^2 =1140$) and with
      oscillation ($\sum \chi^2 = 515$); c) the comparison of $R_{th}$ and $R_{exp}$.
      All three methods show the presence of a periodical structure.
     Data-driving method shows the existence of such a periodical structure in both crossing symmetric reactions
  of   $pp$ and $p\bar{p}$ elastic scattering. It is very important  that the oscillation term has the crossing odd properties.
     Hence, probably, it is part of the Odderon amplitude. Our analysis shows the logarithmic growth of a term
     like this.

  The  corresponding  hadron potential has no simple exponential behavior.
Our   calculations   
show that it has a  modified Gaussian form with probably a cut at large distances.
This may be due to the glueball states of the gluon.
In such a form, the gluon can be distributed at large
distances above the confinement level.
 These new effects are very important  in searching for new physics in the framework of the Standard Model,
for example, checking up the dispersion relations, analyticity and crossing symmetry
of the scattering amplitude.
 This should be especially  considered 
 when determining 
 the total cross sections, the ratio of
the elastic to the total cross sections, and the energy dependence of 
$\rho(s,t)$  (the ratio of the real to the imaginary part of the elastic scattering
amplitude).

      These results allow us to make some predictions for  hadron interactions
      at essentially larger energies at future colliders and at  ultra-high energies of cosmic rays.
        These  effects are likely to exist also in  experimental data
      at essentially smaller energies \cite{osc-conf}
      but  they might have a more complicated form
      (with two different periods, for example).
      Also, these effects impact  our description of spin-dependent hadron interactions
      at low energies (including the energy of NICA project).

\vspace{0.5cm}
{\bf Acknowledgements}
{\small \hspace{0.3cm} OVS would  like to thank O. Teryaev and Yu. Uzikov
for their kind and helpful discussion.
This research was carried out at the expense of the grant of the Russian Science Foundation No. 23-22-00123, \\
 https://rscf.ru/project/23-22-00123. }\\



\vspace{0.5cm}
      {\bf Appendix A: Eikonalization of the scattering amplitude}

   The total elastic amplitude in general gets five helicity  contributions, but at
   high energy it is {sufficient to keep} only spin-non-flip hadron amplitudes.
 The final elastic  hadron scattering amplitude is obtained after the unitarization of the  Born term.
    So, first, we have to calculate the eikonal phase
   \begin{eqnarray}
 \chi(s,b) \   = -\frac{1}{2 \pi}
   \ \int \ d^2 q \ e^{i \vec{b} \cdot \vec{q} } \  F^{\rm Born}_{h}\left(s,q^2\right)\,
 \label{chi}
 \end{eqnarray}
  and then obtain the 
   hadron scattering amplitude 
    \begin{eqnarray}
 F_{h}(s,t) = i s
    \ \int \ b \ J_{0}(b q)  \ \Gamma(s,b)   \ d b\,  \ \ \ \\
 {\rm  with} \ \ \
  \Gamma(s,b)  = 1- \exp[\chi(s,b)].
 \label{Gamma}
\end{eqnarray}
 Numerical integration allows one to calculate an additional term
 in both  the direct and  eikonal approach. \\

{\bf Appendix B: High energy generalized structure model}

    As a basis, we take our HEGS model \cite{HEGS0,HEGS1}
    that quantitatively  describes, with only a few parameters, the   differential cross section of $pp$ and $p\bar{p}$
  from $\sqrt{s} =9 $ GeV up to $13$ TeV, including the Coulomb-hadron interference region and the high-$|t|$ region  up to $|t|=15$ GeV$^2$
  and quantitatively well describes the energy dependence of the form of the diffraction minimum \cite{HEGS-min}.
   However, to avoid  possible problems
 connected with the low-energy region, we consider here only the asymptotic variant of the model \cite{HEGSh}.
   The total elastic amplitude in general receives five helicity  contributions, but at
   high energy it is enough to write it as $F(s,t) =
  F^{h}(s,t)+F^{\rm em}(s,t) e^{i\varphi(s,t)} $\,, where
 $F^{h}(s,t) $ comes from the strong interactions,
 $F^{\rm em}(s,t) $ from the electromagnetic interactions and
 $\varphi(s,t) $
 is the    
   relative phase between the electromagnetic and strong
 interactions   \cite{Can,Petrovphase,PRD-Sum}.    
    The Born term of the elastic hadron amplitude at large energy can be written as
    a sum of two pomeron and  odderon contributions,
    \begin{eqnarray}
 F_{\mathbb{P} }(s,t) & =& \hat{s}^{\epsilon_0}\left(C_{\mathbb{P}} F_1^2(t)  \ \hat s^{\alpha' \ t} + C'_{\mathbb{P}}  A^2(t) \ \hat s^{\alpha' t\over 4} \right), \\
 F_{\mathbb{O} }(s,t) & =&  i \hat{s}^{\epsilon_0+{\alpha' t\over 4}} \left(  
   \pm C'_{\mathbb{O}} t/(1-r_{\mathbb{O} }^{2} t ) \right) A^2(t).
 \end{eqnarray}
 Finally, the Born amplitude is
 $   F_{Born}(s,t) =  F_{\mathbb{P} }(s,t) + F_{\mathbb{O} }(s,t)$.
  The simultaneously fitting procedure of all 24 sets lead to the value of the parameters:
  $C_{\mathbb{P}}=3.30 \pm 0.02$ GeV$^{-2}$, $ C'_{\mathbb{P}}=1.39\pm0.02$ GeV$^{-2}$,
  $C'_{\mathbb{O}}=-0.56\pm0.03$ GeV$^{-2}$, $r_{\mathbb{O}}^2 = 4.9\pm 1.4$ GeV$^{-2}$.
 The sizes  and the energy and momentum transfer
 dependence of the real part of the elastic scattering amplitude $Re F(s,t)_{B}$
  are determined by the complex energy
      $\hat{s}=s   exp(-i\pi/2)$. Hence, the model does not introduce some special functions or assumptions  for  {$Re F_{B}(s,t)$}.
 All terms are supposed to have the same intercept  $\alpha_0=1+\epsilon_0=1.11$, and the pomeron
 slope is fixed at $\alpha'=0.24$ GeV$^{-2}$.
  The model takes into account  two hadron form factors $F_1(t)$ and $A(t)$, which correspond to  the charge and matter
  distributions \cite{GPD-PRD14}. Both form factors are calculated  as the first and second moments of  the same GPDs.
   Taking into account the Mandelstam region of the analyticity of the scattering amplitude
   for the $2 \rightarrow 2 $ scattering process with identical mass
   $s+u+t=4 m_{p}^2$   one takes the normalized energy variable $s$ in complex form $\hat{s}/s_{0}$ with
    $s_{0}=4 m_{p}^{2}$, where
     $m_{p}$ is the mass of the proton.

     In the present model, 
     a  small additional term is introduced into the slope, which reflects some possible small nonlinear
        properties of the intercept.
 As a result, the slope of the amplitude
 in the form    $\hat s^{\alpha'  t} = exp^{B(s,t) t} $
 is taken as
 $$B(s,t) \ =
  \alpha^{\prime} \ln(\hat{s}/s_{0}) (1 -k_1 t/(\ln(\hat{s}/s_0))/k e^{k_2 \ln(\hat{s}/s_{0}) t}).$$
   This form leads to the standard form of the slope as $t \rightarrow 0$ and $t \rightarrow \infty$.
   Note that our additional term at large energies has a similar form as an additional term to the slope
   coming from the $\pi$ loop examined in Ref. \cite{Gribov-Sl} and recently in Ref. \cite{Khoze-Sl}. \\

{\bf Appendix C:  Fitting procedure}

 There are two essentially  different ways of including statistical
 and systematic uncertainties in the fitting procedure \cite{Orava-Sel}.
 The first one, mostly used in connection with the differential cross sections  (for example \cite{Kohara,Pakanoni}),
 takes into account  statistical and systematic errors in quadrature form:
    $\sigma_{i(tot)}^2 = \sigma_{i(stat)}^2+ \sigma_{i(syst)}^2$
   and 
\begin{eqnarray}
\chi^{2}=   \sum_{i=1}^{n} \frac{ ( \hat{E}_{i}  - F_{i}(\vec{a}) )^2  }
{\sigma_{i(tot)}^{2}},
    \label{eq6}
 \end{eqnarray}
 with a standard definitions: $ \hat{E}_{i}$ - experimental data and
 $F_{i}(\vec{a})$ - model calculations with parameters $\vec{a} $.

   The second 
   method accounts for the basic property of systematic uncertainties,
   i.e. the fact that these errors have the same sign and size in proportion to the effect
   in one set of experimental data and possibly  have a different sign and size in another set.
    To account for these properties, extra normalization coefficients
          $k_{j} = 1 \pm \sigma_{j} $ 
    for the measured data
     are introduced in the fit.
     This method is often used by research collaborations to extract,
      for example, the parton distribution functions of nucleons   
        \cite{Stump01,exmp1} and nuclei \cite{EPPS16}
       in high energy accelerator experiments, or in astroparticle physics  \cite{Koh15}.
       In this case, $\sigma_{i(tot)}^2 = \sigma_{i(stat)}^2$ and
   the systematic uncertainty are taken into account as an additional normalization coefficient,
 $k_j$, and the size of $\sigma_j$  is assumed to have a standard systematic error,
   \begin{eqnarray}
\chi^{2} =   \sum_{j=1}^{m}  \left\{  \sum_{i=1}^{n}  \frac{ ( k_{j} \hat{E}_{ij}  -  F_{ij} )^2  }{ \sigma^{2}_{ij(st.)} }  + \frac{(1-k_j)^2}{\sigma^{2}_{j}} \right\}.
\end{eqnarray}
%
   In the first case, the "quadrature form" of the experimental uncertainty gives a wide corridor
   in which different forms of the  theoretical amplitude can exist. In the second case,
   the "corridor of the possibility" is essentially narrow, and it restricts  different forms
   of theoretical amplitudes \cite{Orava-Sel}.  \\

{\bf Appendix D:  Statistical method}

The usual method of minimization $\chi^2$ in this situation often works
 poorly. On the one hand, we should define a certain model for  part of
 the scattering amplitude  having zeros in the domain of small $t$.
 However, this
 model may slightly differ from a real physical picture.
 On the other hand, the
 effect is rather small and gives an insignificant change
 in the sum of $\chi^2$.
 Therefore, in this work let us apply first another method, namely, the method of
 comparing of two statistically independent  choices,
  for example \cite{Hud}.
   If we have two statistically independent choices
  $x^{'}_{n_1}$  and $x^{"}_{n_2}$
  of values of the quantity  $X$  distributed around
  a definite value of $A$ with the standard error  equal to $1$,
  we can try
  to find the difference between
  $x^{'}_{n_1}$  and $x^{"}_{n_2}$.
  For that we can compare
  the arithmetic mean of these choices
$ \dd X = (x^{'}_1 + x^{'}_2 + ... x^{'}_{n1})/n_1  -
        (x^{"}_1 + x^{"}_2 + ... x^{"}_{n2})/n_2  =
     \overline{x^{'}_{n_1}} - \overline{x^{"}_{n_2}}.   $
  The standard deviation for that case will be
$   \delta_{\overline{x}} = [1/n_1 +1/n_2]^{1/2} $.
 And if $\dd X / \delta_{\overline{x}}$ is larger than $3$, we can say
 that the difference between these two choices has  the
  $99\%$ probability.


\begin{thebibliography}{99}


\bibitem{Block-85} Block M.M., Cahn R.N., Rev.Mod.Phys., {\bf 57},
563 (1985).
%
\bibitem{royt}  Roy S.M., {\it Phys.Lett.} {\bf B 34}, 407 (1971).

 \bibitem{Bogolyubov:1983gp}
 {Bogolyubov N.N., Shirkov D.V.} {Quantum Fields}. 
 \newblock Benjamin-Cummings Pub. Co, (1982).
\bibitem{Drem1} 
  I.M. Dremin, Int. J. Mod. Phys. A, {\bf 31}, 1650107  (2016).
\bibitem{akm} Auberson G., Kinoshita T., Martin A., {\it Phys. Rev.}
            {\bf D3}, 3185 (1971).


\bibitem{zar} Starkov   N.I.,  Tzarev  V.A.,
     Lett. JTEPh, {\bf 23}, 403 (1976).




\bibitem{barsh} Barshay S.,  Heiliger P., {\it Z. Phys.} {\bf C 64} 675 (1994).

\bibitem{kontr} Kontros J., Lengyel A.,  Ukr.J.Phys. {\bf 41},  290 (1996).


\bibitem{CS-PRL} Cudell J.R., Selyugin O.V., Phys.Rev.Lett. {\bf 102}, 032003 (2009).

\bibitem{L-range} Cudell J.-R.,  Selyugin O.V., Aip Conf.Proc. 1350: 115 (2011);
arxiv: 0811.4369; arxiv:1011.4177.


\bi{selh95} O.V. Selyugin ,
     Ukr.J.Phys. {\bf 41}, 296 (1996).

\bibitem{gnsosc} Gauron P., Nicolescu B., Selyugin O.V.,
      Phys.Lett, {\bf B 397}, 305 (1997).


\bibitem{Per-1} P. Grafstrom, arXiv: 2307.15445.


  \bibitem{Jenk-rev} L. Jenkovszky, R. Orava, E. Predazzi, A. Prokudin, O.Selyugin,
   Mod.Phys. A {\bf 24}, 2551 (2009).






 \bibitem{Pakanoni} 
 G.Pancheri, S. Paccetti, V. Strivasta, Phys.Rev. D, {\bf 99}, 034014  (2019).



 \bibitem{Soffer-Wu} C. Bouraly, J.Soffer, T.T. Wu, Eur.Phys.J. C {\bf 28}, 97 (2003).

        \bibitem{Bourrely-14} C. Bourrely, Eur.Phys.J. C {\bf 74}, 2736 (2014).


     \bibitem{Kohara} E. Ferreira, A. K. Kohara, J. Sesma,
     Phys. Rev. C, {\bf 97} 014003 (2018).


 \bibitem{HEGS1} Selyugin O.V.,
  Phys.\ Rev.\ D {\bf 91}, no. 11, 113003 (2015).







   \bibitem{Martynov}  E. Martynov, B. Nicolescu, Eur. Phys. J. C, {\bf 56}, 57
(2008).

   \bibitem{Petrov-Tkach} V. Petrov, N. Tkachenko,
       Phys. Rev. D  {\bf 106}, 054003 (2022).






 \bibitem{HEGS0} Selyugin O.V.,
  Eur.\ Phys.\ J.\ C {\bf 72}, 2073 (2012).




 \bibitem{Muller}
M{\"u}ller et al.,
Fortschr. Phys. , {\bf 42}, 101 (1994).


\bibitem{Ji97}
{Ji, X.}
 Phys. Rev., {\bf D55}, 7114 (1997).


\bibitem{R97}
 Radyushkin, A.V.,
 Phys. Lett. B, {\bf 380}, 41 (1996).


\bibitem{Diehl}
{Diehl, M.}
Eur. Phys. J. C, {\bf 25}, 223 \textbf{2002}.



\bibitem{GPD-ST-PRD09}  O.V. Selyugin and O.V. Teryaev,
 Phys.Rev. D {\bf 79}, 033003 (2009). 







\bibitem{GPD-PRD14}   Selyugin  O.V.,
        Phys. Rev. D, {\bf 89}, 093007  (2014). 






\bibitem{data-Sp}
http://durpdg.dur.ac.uk/hepdata/reac.html.

\bibitem{Land-Bron} K.R. Schubert, In Landolt-Bronstein, New Series, v. 1/9a, (1979).



 \bibitem{Khoze-Sl} Khoze,  V.A.; Martin, A.D.; Ryskin, M.G.
 J. Phys. G, {\bf 42}, 025003  (2015) .


 \bibitem{T7a}  G. Antchev et al. (TOTEM Coll.)
  Eur.Phys.Lett., {\bf 95} 41001 (2011).



\bibitem{T2p76}  Antchev G.  TOTEM Collaboration; et al.
 Eur. Phys. J. C, {\bf 80}, 91 (2020).


\bibitem{64-T8}
The TOTEM Collaboration (G. Antchev et al.)
       Nucl. Phys. B {\bf 899},  527  (2015).


     \bibitem{TOTEM13-1set} Antchev G. {\it et al.} [TOTEM Collaboration],
  arXiv:1712.06153 [hep-ex].

\bibitem{TOTEM13-2set}  Antchev  G.{\it et al.} [TOTEM Collaboration],
  arXiv:1812.08283 [hep-ex].




  \bibitem{L-P-22} 
G.B. Bopsin, E.G.S. Luna, A.A. Natale, and M. Peláez,
Phys. Rev. D {\bf 107}, 114011 (2023).
\bibitem{ATLAS-13} ATLAS Collaboration, hep-ex: 2207.12246.

 \bibitem{70-T8c} G. Antchev (TOTEM Collaboration),Eur,Phys.J. C {\bf 82} 263  (2022).

  \bibitem{TOTEM-13a} G. Antchev (TOTEM Collaboration),Eur,Phys.J. C {\bf 79}, 861 (2019).

  \bibitem{TOTEM-13b} G. Antchev (TOTEM Collaboration),  hep-ex:2111.11991.




 \bibitem{65-T8b} G. Antchev (TOTEM Collaboration),
 Eur. Phys. J. C {\bf 76}, 661  (2016).


  \bibitem{TOTEM-8nexp} The TOTEM Collaboration (G. Antchev et al.),
  Nucl. Phys. B, {\bf 899}, 527 (2015).

\bibitem{63-ATLAS-8} ATLAS Collaboration,
Phys. Lett. B {\bf 761}  158 (2016)














\bibitem{D0-1p96}V.M. Abazov et all.  (D0 Collaboration),
 Phys.Rev. D {\bf 86},  051502  (2012).




\bibitem{E710-1p8} N. Amos et all. (E710 Collaboration),
 Phys.Rev.Lett. {\bf 68}  2433  (1992).


 \bibitem{CDF-1p8a} F. Abe et all. (CDF Collaboration)
   Phys.Rev. D, {bf 50},  5518  (1993).

   \bibitem{Bernard86}  D. Bernard et all. (UA4 Collaboration),
 Phys. Lett. B {\bf 171},  142  (1986).



 \bibitem{UA4/2}  C. Augier et all., UA4/2 Collaboration,
 Phys.Lett.B {\bf 316}, 448 (1993).

   \bibitem{UA4-87} D. Bernard et all. (UA4 Collaboration),
 Phys. Lett. B {\bf 198},  583  (1987).




 \bibitem{UA4-87-21} M. Bozzo et all. 
  Phys. Lett. B {\bf 147}, 385 (1984);



 \bibitem{UA4-87-22} R. Batinston et all. 
  Phys. Lett. B {\bf 127}, 472 (1983);


 \bibitem{UA4-87-23} M. Bozzo et all. et all. 
  Phys. Lett. B, {\bf 155}, 197 (1985);
   CERN-EP-85-31
















  \bibitem{osc-13} Selyugin O.V.
\emph{ Phys. Lett. B }{\bf 2019} \emph{797}, \mbox{134870--134873.}


 \bibitem{Hud}  Hudson D.J., STATISTIC, Lectures on Elementary Statistics and Probability, Geneva (1964).




\bibitem{osc-3p7}  O.~V.~Selyugin, J.-R. Cudell, Mod.Phys.Lett. A {\bf 27}, 1250113  (2012);
 arxiv: 1207.0600.





\bibitem{fd13} Selyugin O.V.
Mod. Phys. Lett. A, {\bf 36},  2150148-1 (2021).





\bibitem{fd-LHC} Selyugin O.V.
Symmetry, {\bf 15}, 760  (2023).

  \bibitem{Her} V.M. Abazov et all.,
   Phys.Rev.Lett, {\bf 127},  062003 (2021).  D0-ATLAS




   \bibitem{Sitnik1} I.M. Sitnik,
 Comp.Phys.Comm., {\bf 185}, 599 (2014).
%
%

\bibitem{Sitnik2}  I.M. Sitnik, I.I. Alexeev, O.V. Selyugin,
 Comp.Phys.Comm., {\bf 251}, 107202 (2020).


\bibitem{Gribov-Sl} A. Anselm and V. Gribov,
  Phys.\ Lett.\  B, {\bf 40}, 487 (1972). 

\bibitem{HEGS-min} Selyugin O.V.,
  Nucl.\ Phys.\ A {\bf 959}  116  (2017).



 \bibitem{HEGSh} Selyugin O.V.,  Cudell J.-R., arxiv: 1810.11538




\bibitem{osc-conf} Selyugin O.V.,  Cudell J.-R.,
Mod.Phys.Lett. A, {\bf 27}, 1250113  (2012).















%
%























\bibitem{Can}       R. Cahn,
      Zeitschr. fur Phys. C, {\bf 15}, 253 (1982). 


 \bibitem{Petrovphase} V.A. Petrov,
Proceedings of the Steklov Institute of Mathematics, {\bf 309}, 219 (2020). 

\bibitem{PRD-Sum}
  O.\,V.~Selyugin,
  Phys.\ Rev. D,  {\bf 60}, 074028  (1999).









   \bibitem{Orava-Sel} R. Orava, O. V. Selyugin, arxiv: 1804.05201.


\bibitem{Stump01} D. Stump, et all. 
  Phys.ReV. D, {\bf 65}, 014012 (2001). 

 \bibitem{exmp1}  P. Jimenez-Delgado, Phys.Lett. B, {\bf 714} 301 (2012).

     \bibitem{EPPS16} K. J. Eskola et al., Eur.Phys.J. C, {\bf 77}, 163  (2017).

 \bibitem{Koh15} F. Kohlinger, H. Hoekstra, and M. Eriksen,
 Mon Not R Astron Soc, {\bf 453}, 3107 (2015).




\end{thebibliography}
\end{document}